# The NIKA instrument: results and perspectives towards a permanent KID based camera for the Pico Veleta observatory

A. D'Addabbo[1], R. Adam[2], A. Adane[3], P.Ade[4], P. André[5], A. Beelen[6], B. Belier[7], A. Benoit[1], A. Bideaud[4], N. Billot[8], O.Bourrion[2], M. Calvo[1], A. Catalano[2], G. Coiffard[3], B. Comis[2], F.-X. Désert[9], S. Doyle[4], J. Goupy[1], C. Kramer[8], S. Leclercq[3], J. Macias-Perez[2], J. Martino[6], P. Mauskopf[4,10], F. Mayet[2], A. Monfardini[1], F. Pajot[6], E. Pascale[4],  N. Ponthieu[9], V. Revéret[5], L. Rodriguez[5], G. Savini[4,11], K. F. Schuster[3], A. Sievers[8], C. Tucker[4], R. Zylka[3]

*Abstract* — **The New IRAM KIDs Array (NIKA) is a pathfinder instrument devoted to millimetric astronomy. In 2009 it was the first multiplexed KID camera on the sky; currently it is installed at the focal plane of the IRAM 30-meters telescope at Pico Veleta (Spain). We present preliminary data from the last observational run and the ongoing developments devoted to the next NIKA-2 kilopixels camera, to be commissioned in 2015. We also report on the latest laboratory measurements, and recent improvements in detector cosmetics and read-out electronics. Furthermore, we describe a new acquisition strategy allowing us to improve the photometric accuracy, and the related automatic tuning procedure.**

I. INTRODUCTION

KINETIC INDUCTANCE DETECTORS (KIDs) have recently drawn the attention of the low-temperature detectors community. Reduced fabrication complexity, high sensitivity, small time constants and most notably the intrinsic capability to frequency multiplexed readout open new possibilities to applications that need very large array size and/or high speed read-out [1] . Lumped Element Kinetic Inductance Detectors (LEKIDs) [2] designed and fabricated in our collaboration have already shown good on-sky performance as demonstrated by recent NIKA observing runs.

The NIKA camera has been developed in Grenoble to work at the focal plane of the IRAM (Institut de Radioastronomie Millimétrique) 30-m telescope at Pico Veleta. In its latest configuration, the instrument consists of a dual-band camera with band centered at 150 GHz and 240 GHz, equipped with respectively 132 and 224 pixels based on LEKIDs. The first four NIKA commissioning campaigns at IRAM 30-m telescope [3] [4] demonstrated performances comparable to state-of-the-art bolometer arrays, operating at the same wavelengths [5]. Since 2012, the NIKA camera the NIKA camera has been permanently installed at the IRAM 30-m telescope.

We report here on the latest improvements of the instrument adopted during the last two observational runs, carried out in October 2011 (3$^{rd}$ run) and November 2012 (5$^{th}$ run). We present the new HILBERT pattern for LEKID detectors, allowing us to increase the quantum efficiency of LEKIDs. We briefly describe our read-out electronics, focusing on the new modulated calibration method. We also describe the two main improvements in the acquisition strategy introduced in the last run: the absolute sky dip calibration and the real time auto-tuning. We conclude by presenting our results from the last observational run, including the sensitivity, the resolution and the cosmetics of the focal plane of the NIKA instrument.

II. THE NIKA CAMERA

The IRAM 30 meters telescope is located in a dry area at 3000m a.s.l. on Pico Veleta, Spain. With a Nasmyth  field of view of 6.5 arcmin and its remarkable angular resolution, it is among the most powerful telescopes in the millimeter band. In order to completely exploit its potential, large arrays of thousands of detectors are needed. Full-sampling arrays with up to thousands of bolometers are reaching maturity, but further array scaling is limited by the multiplexing factor of the read-out electronics. For this reason it is worth developing

[1] Institut Néel, CNRS and Université de Grenoble, France
[2] Laboratoire de Physique Subatomique et de Cosmologie (LPSC), CNRS and Université de Grenoble, France
[3] Institut de RadioAstronomie Millimétrique (IRAM), Grenoble, France
[4] Astronomy Instrumentation Group, University of Cardiff, UK
[5] Laboratoire AIM, CEA/IRFU, CNRS/INSU, Université Paris Diderot, CEA-Saclay, 91191 Gif-Sur-Yvette, France
[6] Institut d'Astrophysique Spatiale (IAS), CNRS and Université Paris Sud, Orsay, France
[7] Institut d'Electronique Fondamentale (IEF), Université Paris Sud, Orsay, France
[8] Instituto Radioastronomia Milimétrica (IRAM), Granada, Spain
[9] Institut de Planétologie et d'Astrophysique de Grenoble (IPAG), CNRS and Université de Grenoble, France
[10] SESE and Dept. of Physics, Arizona State University, Tempe, AZ, USA
[11] University College London (UCL), UK

E-mail: antonio.d-addabbo@grenoble.cnrs.fr

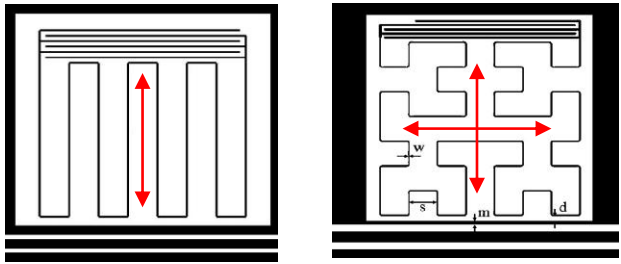

Fig. 1. On the left, standard LEKID design, sensitive only to one polarization. On the right, Hilbert LEKID design, sensitive to the two polarization.

a camera based on an alternative technology that does not suffer from this limitation. Thanks to their intrinsic capability of frequency domain multiplexing, Kinetic Inductance Detectors provide a very effective solution.

The NIKA camera is a dual-band KIDs camera custom designed for the IRAM 30-m Nasmyth-focus telescope. It consists of two arrays of KIDs, cooled down to about 70 mK with a 4He-3He dilution cryostat. The two arrays have their maximum of sensitivity at about 140 GHz and 240 GHz, with respective angular resolution (FWHM) of 18.5 arcsec and 12.5 arcsec, the effective fields of view are about 2.1 and 1.7 arcmin in diameter. For the first four runs, the number of detectors was the same, 132, for the two different bands. For the fifth run in November, 2012, the 240 GHz channel was increased to 224 pixels.

### III. HILBERT LEKIDs

In the 3$^{rd}$ observational run on October 2011 a new type of LEKID was adopted: the Hilbert LEKID. It consists in a lumped element KID in which the classical meander geometry has been replaced by a Hilbert-like fractal pattern of the 3$^{rd}$ order [6]. The Hilbert LEKID has an absorbing area uniformly filled by the pattern shown in fig. 1, that allows the detector to be sensitive to the two polarizations, with no preferential direction in absorption (see fig. 2).

Maintaining the same performances for each of the two polarizations, a factor 2 in quantum efficiency is readily achievable by using a dichroic instead of a wire grid polarizer to split the beam.

### IV. READ-OUT AND ACQUISITION STRATEGY

From October 2011, we introduced a new relative photometric calibration that allowed us to improve its accuracy by a factor 3. To better understand how it works and what actually are the main improvements with respect to the old acquisition strategy, a quick overview of the read-out electronics system is given.

In order to monitor the signal of kilo-pixels arrays, the NIKA readout electronics adopts the Frequency Domain Multiplexing approach typical of KIDs arrays [7]. Briefly, each electronic board generates the two frequency combs (each tone phase shifted by 90° between I and Q), which are then up-converted to the frequency band of our resonators by mixing them to a Local Oscillator (LO) signal. The resulting comb of tones is fed to a programmable attenuator for power adjustment. After passing through the cryostat and the low noise amplifier, the signal is down-converted back to the baseband. Then it is acquired using a fast ADC by comparing this signal to the reference one sent at the cryostat input (see fig. 3). Acquiring for each resonance the (I, Q) vector, we can read-out the signal both in amplitude and in phase. We decide to use phase instead of amplitude, because of its higher responsivity. We then monitor the on-resonance phases of each pixel.

In order to convert the raw data to sky signal, we need to calibrate the change in power for a given change in phase. Because the shift in resonance frequency of the KIDs is proportional to its incident power load [8], we only need a relation that link the shift in resonance frequency ($\Delta f$) and the shift in phase ($\Delta \phi$) (relative photometric calibration). A calibration source on the sky then provides us with absolute photometric calibration.

*A. Relative photometric calibration: the modulated read-out*

We describe here the two different methods we have used along the campaign to calibrate the on-resonances shifts in phase.

The old approach was to characterize the response of the detectors *before* the observations. Thanks to a frequency sweep around the resonances, we fitted the center of the typical IQ resonance circles to recalculate the phase with respect to the fitted center (see fig. 4), so that the shift in phase is linear with the incident power [9]. This sort of off-line calibration suffers from changes in background conditions. Resonance circles rapidly change with the power load, shifting the optimal working point and lowering photometric accuracy, so that many sweeps are needed to recalibrate the resonances.

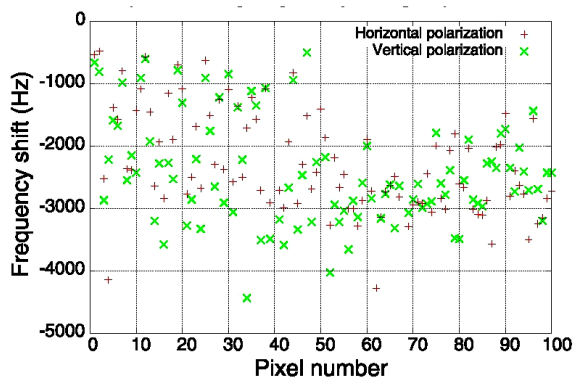

Fig. 2. Comparison of the two polarization response for an Hilbert LEKID array. For each pixel of the array the response in resonance frequency shift are plotted

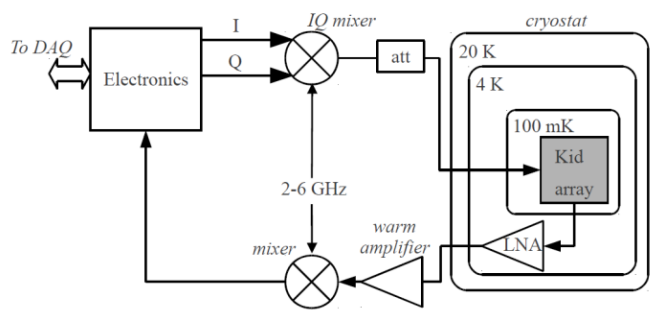

Fig. 3. A schematic overview of the NIKA setup for an electronic board. By using *IQ mixers*, the LO signal at GHz frequencies *(2-6 GHz)* is used to up/down convert the *IQ* comb at the input/output of the cryostat. The signal is adjusted in power, fed to the KID array and read after the amplifiers.

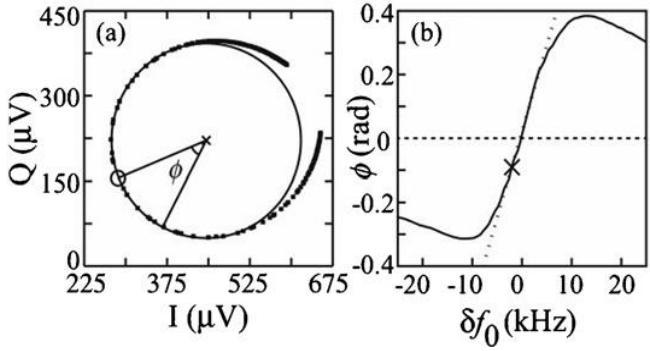

Fig. 4. KID calibration (old approach). On the left, *I–Q* frequency sweep across a resonance. The complex phase is calculated by fitting each resonance to a circle (shown); the phase angle is then determined with respect to the center of the resonance circle as indicated by the ×. The minimum transmission is marked with a small circle. On the right, frequency shift vs. complex phase. The × on the curve indicates the maximum frequency shift when Mars transits the pixel

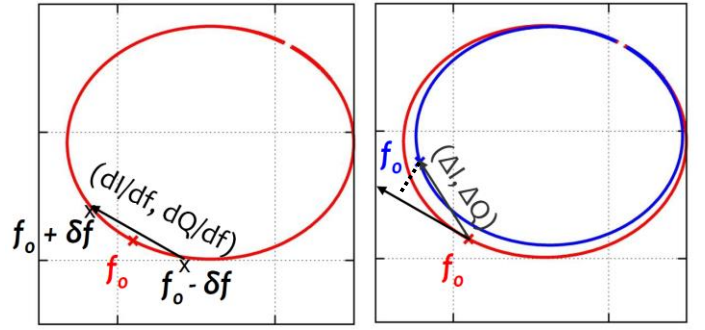

Fig. 5. Modulated read-out. On the left, a classical sweep in frequency around a resonance (red circle) in the *IQ* plane, $f_0$ is the resonance frequency. The two points in frequency $f_0 + \delta f$ and $f_0 - \delta f$ are indicated by the × on *IQ* circle. On the right, *IQ* circle (blue) changing with an incoming power. The gradient vector (black) is tangent to the resonance circle in the working point. It is used to obtain the shift in frequency simply projecting the vector $(\Delta I, \Delta Q)$ between two successive observations (blue vector) on this gradient.

To avoid these drawbacks, from the 3$^{rd}$ run on, a new approach has been adopted. We modulate the LO carrier frequency, evaluating for each point not only *(I,Q)*, but also *(dI/df, dQ/df)*, see fig. 5. This provides us with a "reference gradient" that can be used to calibrate in frequency the shifts of the resonance. This can be done by calculating the vector $(\Delta I, \Delta Q)$ between two successive observations, and projecting it on the reference gradient. The shift in frequency $\Delta f$ is given by

$$\Delta f = \frac{(\Delta I, \Delta Q) \cdot (dI/df, dQ/df)}{(dI/df, dQ/df)^2} \cdot \delta f_{LO} \quad (1)$$

where $\delta f_{LO}$ is the amplitude of the LO modulation (see fig. 5).

The main advantage of the modulated read-out is that the reference gradient vector is automatically updated as the observing conditions change, obtaining a sort of *real time* calibration. This new calibration method leads to an improvement by a factor 3 in terms of relative photometric accuracy compared to the previous run [4].

*B. Absolute photometric calibration: the sky dip*

In order to estimate the incoming flux from an astrophysical source, another crucial point is to correct for the contribution of the atmosphere. We can estimate its opacity $\tau$ starting from the KIDs resonances themselves. The idea is that as the atmospheric opacity increases, the resonance frequencies diminish due to the increasing background load on the detectors. In principle we are thus able to estimate $\tau$ by calculating the mean value of all the resonance frequencies $<f_0^i>$. To calibrate the relation $\tau = \tau(<f_0^i>)$ we first used the resident tau-meter at the Pico Veleta observatory.

A new approach, not requiring the tau-meter anymore, consists in a sky dip with the telescope, in order to trace the *elevation* vs. $<f_0^i>$ curve and then evaluate $\tau$ by its fit. This provides us with an integrated tau-meter and a powerful self-consistent strategy to evaluate atmospheric opacity, that brings absolute photometric accuracy to around the 10% (see fig. 6).

V. AUTO-TUNING

In this paragraph we focus on the strategy to adjust the resonance frequencies during the observations. As already said, the observing conditions often change rapidly. These changes modify the optimal working point, so that a sweep in frequency is required at the end of each observation, to tune again the frequencies. This wastes a lot of time at the expense of scientific measures. Our auto-tuning procedure is an algorithm that quantifies the shift in frequency due to background fluctuations, without the need for a frequency sweep.

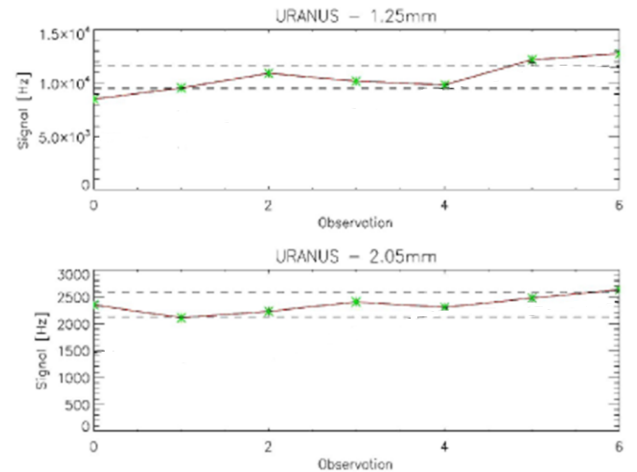

Fig. 6. Absolute photometric calibration. Here is plotted the measured flux from Uranus, given in resonance frequency shift units, for the two channels of the NIKA camera. The sky dip method shows a dispersion of about 10% between measurements from different observations.

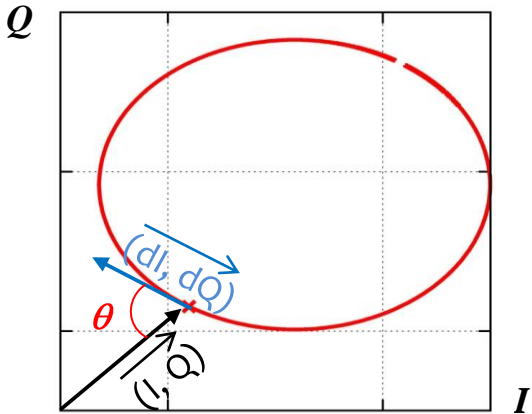

Fig. 7. Definition of the angle as the angle the vectors *(I, Q)* and *(dI/df, dQ/df)* in the IQ plane.

The auto-tuning uses the angle $\theta$ between the vectors *(I, Q)* and *(dI/df, dQ/df)* to characterize the resonance (see fig. 7). The *tuning curve* $\theta$ vs. $f$ is obtained by the initial frequency sweep. First of all, we rescale the $\theta$ angle so that $\theta = 0$ on the resonance. Suppose now that after some time we find $\theta \neq 0$. This means that the resonance shape has changed and its optimal frequency has shifted. We are actually on another $\theta$ vs. $f$ tuning curve, the one corresponding to the present background conditions (see fig. 8). The auto-tuning algorithm evaluates the shift in frequency as we still were on the last tuning curve and re-tune applying this correction. The method works properly (converges) if the changes in background are not so strong, in such a way that we remain in the linear region of the tuning curve. We can iterate the process as many times as we want. A single iteration of the auto-tuning process only takes a few seconds, so that we are able to complete a full tuning cycle in a dozen of seconds, saving a lot of observational time.

## VI. LAST RESULTS

During the scientific run on November 2012, the 140 GHz channel was used with 127 (on 132) detectors with mean effective sensitivity of 15 mJy s$^{1/2}$ per beam, while 91 detectors with mean effective sensitivity of 81 mJy s$^{1/2}$ per beam were used for the 240 GHz channel. The latter sensitivity was due to a malfunction in a cold amplifier during the campaign. Using only the 8 central detectors of this array, the expected mean effective sensitivity of 32 mJy s$^{1/2}$ per beam is recovered. The fig. 9 shows the geometry of the two arrays.

## VII. CONCLUSION

The NIKA camera has demonstrated the capability of multiplexed LEKIDs arrays to take significant data from astrophysical sources. The sensitivities achieved by the NIKA camera are comparable to performances of the state-of-art of bolometers. They demonstrated the potential on the next generation NIKA-2 kilo-pixels camera. NIKA-2 will have 1000 detectors at 140 GHz and 2x2000 detectors at 240 GHz, providing in that band also measurement of the linear polarization. The NIKA-2 instrument, to be commissioned in 2015, will provide a next generation mm-wave facility for astronomical and cosmological observations with remarkable sensitivity, mapping speed and angular resolution.

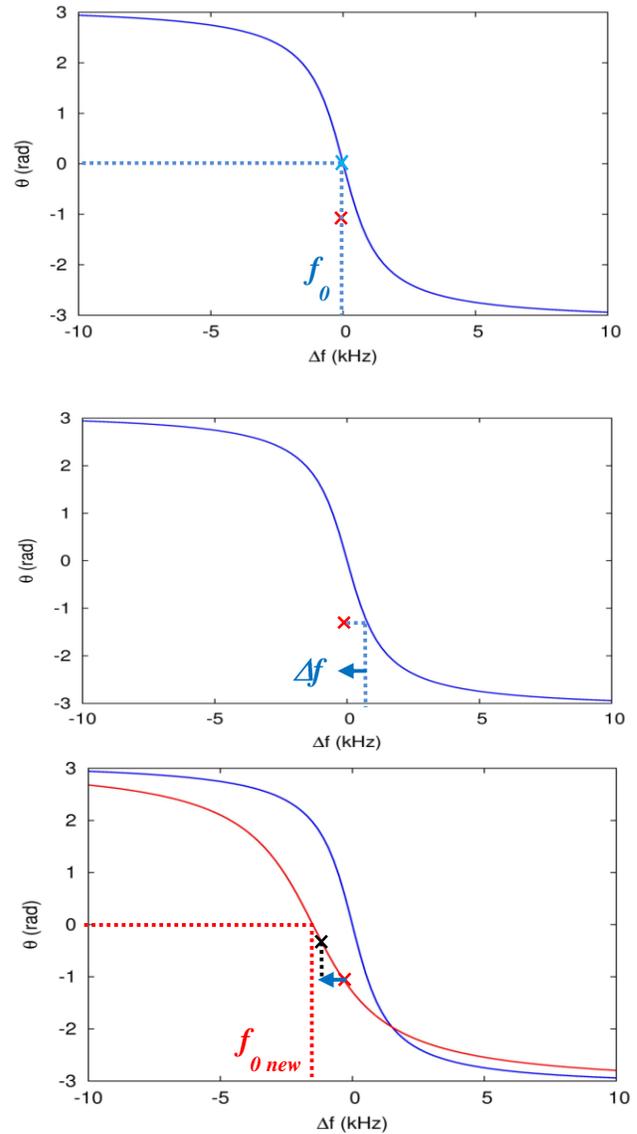

Fig. 8. Real time auto-tuning. All the plots show: on the y-axis the angle $\theta$ between the vectors *(I, Q)* and *(dI/df, dQ/df)*, rescaled to 0 on the resonance; on the x-axis the shift in frequency $\Delta f$ with respect to the initial frequency f$_0$. (Top) At the end of the initial sweep, we are in the working point indicated by a blue ×, that is at $f_0$, moving on the blue tuning curve. As the observational conditions change (but the frequency we use to feed the resonance) the working point shifts to the red ×, because the tuning curve has changed to the red one shown on the bottom plot. As the exact form of the red tuning curve is unknown, (Middle) we calculate the shift in frequency that would be needed to reset the $\Delta f$ if we still were on the blue tuning curve. (Bottom) Applying this correction to f$_0$ we are actually moving on the new tuning curve (the red one). At the end we are on the working point indicated by the black × and then we have approached the new resonance frequency f$_{0\ new}$. We can finally use these last wo points (black and red ×) to evaluate the slope of the new tuning curve and iterate the process.

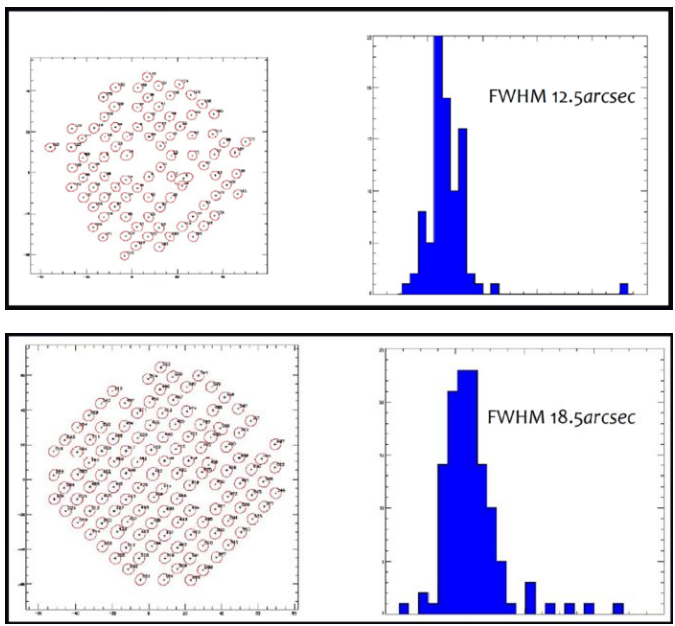

Fig. 9. Optical properties of the two NIKA LEKIDs arrays at 240 GHz (top) and 140 GHZ (bottom). On the left is shown the image of the focal planes. Each point represents the position of one of the functioning pixel. On the right, the histogram of the Full Width Half Maximum value for the beams of the detectors, that determines the angular resolution of the instrument.

ACKNOWLEDGMENT

This work has been partially funded by the Foundation Nanoscience Grenoble, the ANR under the contracts "MKIDS" and "NIKA". This work has been partially supported by the LabEx FOCUS ANR-11-LABX-0013. The collaboration has further benefited from the support of the European Research Council Advanced Grant ORISTARS under the European Union's Seventh Framework Programme (Grant Agreement no. 291294). The NIKA cryostat has been designed and fabricated at the Institut Néel. We acknowledge the crucial contribution of the Cryogenics Group, and in particular Gregory Garde, Henri Rodenas, Jean Paul Leggeri, Philippe Camus.